\documentclass{article}

\begin{document}
\title{On the Reality of Minkowski Space}
\author{Vesselin Petkov \\
Science College, Concordia University\\
1455 De Maisonneuve Boulevard West\\
Montreal, Quebec, Canada H3G 1M8\\
E-mail: vpetkov@alcor.concordia.ca}
\date{}
\maketitle

\begin{abstract}
Should physicists deal with the question of the reality of Minkowski space (or any relativistic spacetime)?
It is argued that they should since this is a question about the dimensionality of the world at the
macroscopic level and it is physics that should answer it.
\end{abstract}

\vspace{.5cm}

Almost a hundred years have passed since 1908 when Hermann Minkowski gave a four-dimensional
formulation of special relativity according to which space and time are united into an inseparable
four-dimensional entity -- now called Minkowski space or simply spacetime -- and macroscopic bodies
are represented by four-dimensional worldtubes. But so far physicists have not addressed the question
of the reality of these worldtubes and spacetime itself since they appear to assume that spacetime and
the worldtubes of physical bodies do not have counterparts in the external world. The reason for such an
assumption is the fact that special relativity can be equally formulated in a three-dimensional and a
four-dimensional language. However, while the two representations of relativity are equivalent in a sense
that they correctly describe the relativistic effects, they are diametrically different in terms of the
dimensionality of the world. As the world is either three-dimensional or four-dimensional only one of the
representations of relativity is correct since only one of them adequately represents the world's
dimensionality at the macroscopic level.

For this reason it is a valid question to ask whether spacetime and the worldtubes of bodies represent
something real or they are just theoretical entities introduced to describe more concisely the relativistic
effects that take place in a three-dimensional world. Put another way, the question is: ``Are the
relativistic effects manifestations of the four-dimensionality of the world as Minkowski advocated?'' Some
physicists appear to think that these are philosophical questions which they should not bother to answer.
However, I doubt whether anyone would seriously argue that the dimensionality of the world is a
philosophical question. Minkowski would certainly not. And I think we owe him, especially now -- a
hundred years after his profound insight -- a thorough examination of his main contribution to physics --
the concept of spacetime.

Dealing with the issue of the reality of spacetime and worldtubes is important for several reasons. It is
natural to resolve this issue, i.e. the question of the dimensionality of the world according to relativity,
first before dealing with the multi-dimensional spaces of more recent theories\footnote{Physicists freely
talk about extra dimensions and parallel universes but do not seem bothered by the fact that the issue of
the nature of the first extra-dimensional world -- spacetime -- remains unresolved. A natural forum where
this issue can be tackled can be the third spacetime conference, organized by the International Society
for the Advanced Study of Spacetime (http://www.spacetimesociety.org/) since it will commemorate the
one hundredth anniversary of Minkowski's talk ``Space and Time''.}. Addressing the nature of spacetime
will allow us to answer the question ``Is relativity possible in a three-dimensional world?'' which will
provide us with a deeper understanding of what this theory tells us about the world. In particular, the
question of the reality of worldtubes, as will be shown below, has immediate implications for the proper
understanding of the physical meaning of the relativistic effects. And that is essential since there still
exist misconceptions about the interpretation of some of these effects. For example, the temptation to
interpret the relativistic length contraction in terms of deformation forces is sometimes difficult to resist
despite the fact that the muon experiment, for instance, completely ruled out such an interpretation by
demonstrating that space itself (where there are obviously no deformation forces) contracts
relativistically as well \cite{ellis}.

A clear view on the nature of spacetime makes it possible to teach relativity far more efficiently.
Everyone who has taught classes on special relativity knows that students often ask questions which
some professors find difficult to answer. For instance, you derive length contraction and try to explain it
with the help of a spacetime diagram which depicts how the instantaneous three-dimensional spaces of
two observers in relative motion intersect the worldtube of a meter stick at different angles. The usual
questions are: ``The two three-dimensional cross-sections of the meter stick's worldtube nicely explain
the effect but is this a true explanation? Is the worldtube of the meter stick a real four-dimensional
object? Is spacetime nothing more than a four-dimensional mathematical space or is it a mathematical
model of a real four-dimensional world with time entirely given as the fourth dimension?''

I think the challenge posed by such questions should be faced. These are deep questions and avoiding
them would hardly be the best scientific and pedagogical approach. The difficulty with these questions is
that their answers, though straightforward, are counterintuitive. In the case of length contraction relativity
of simultaneity clearly demonstrates that while measuring the same meter stick two observers in relative
motion measure two different three-dimensional objects. This becomes completely evident when it is
taken into account that the meter stick (as an extended three-dimensional body) is defined as ``all its
parts which exist \textit{simultaneously} at a given moment of time''. As the two observers have different
sets of simultaneous events it follows that two different three-dimensional meter sticks (two different sets
of simultaneously existing parts of the meter stick) exist for them. Therefore the two observers do
measure two different three-dimensional meter sticks. However, this is possible only if the worldtube of
the meter stick is a real four-dimensional object, which makes it possible for the two observers to regard
two different three-dimensional cross-sections of it as their three-dimensional meter sticks; what is `the
same meter stick' is the meter stick's worldtube. Hence the physical meaning of length contraction turns
out to be profound as Minkowski argued \cite[p. 83]{minkowski} -- length contraction (and therefore
relativity of simultaneity as well) is a manifestation of the four-dimensionality of the world. And indeed, if
the world were three-dimensional, this effect would be impossible -- the meter stick's worldtube would not
be real and therefore both observers would measure the same three-dimensional meter stick (the same
set of simultaneously existing parts of the meter stick), which would mean that the observers would have
a common class of simultaneous events in contradiction with relativity.

Although length contraction alone is sufficient to settle the issue of the dimensionality of the world
consider the following more general argument as well. The world cannot be three-dimensional since such
a world is defined in terms of (i) the pre-relativistic division of events into past, present and future, and
(ii) the pre-relativistic concept of absolute simultaneity -- as everything that exists \textit{simultaneously}
at a given moment. So a three-dimensional world is defined as the class of simultaneous events at a
given moment. To see that the world cannot be three-dimensional according to relativity, assume for a
moment the pre-relativistic view of reality -- that reality is indeed a three-dimensional world. Then it
inescapably follows from relativity of simultaneity that two observers in relative motion have two different
three-dimensional worlds since the observers have different classes of simultaneous events. But this is
possible only if these worlds are different three-dimensional cross-sections of a real four-dimensional
world. As in the case of length contraction it is again evident that relativity of simultaneity is a
manifestation of the four-dimensionality of the world\footnote{As ``in special relativity, the causal
structure of space-time defines a notion of a `light cone' of an event, but does not define a notion of
simultaneity'' \cite{wald2006} one might object against the use of relativity of simultaneity as an argument
to determine the dimensionality of the world. However, such an objection leads to the four-dimensionality
of the world even faster. If reality cannot be described in terms of simultaneity, then the four-dimensional
area lying outside the light cone at an event $P$ should represent what exists. On this view all events in
the past and the future light cone should be regarded as non-existent. By the same criterion, however,
the events of the past and future light cone should also exist since they fall in the area lying outside the
light cone associated with another event $Q$ which is space-like separated from $P$ \cite{weingard}.
Therefore all events of spacetime are equally existent, which means that the world is four-dimensional.}.
That is why no relativity of simultaneity would be possible in a three-dimensional world -- if the world were
three-dimensional, the class of simultaneous events constituting such a world would be common to all
observers in relative motion, which would mean that simultaneity is absolute.

Addressing the question of the nature of spacetime may even pave the way for new results as Minkowski
anticipated \cite[p. 91]{minkowski}. The realization that the worldtubes of physical bodies are real
four-dimensional objects makes it possible to pursue more rigorously his program -- physical laws might
find their most perfect expression as reciprocal relations between worldlines \cite[p. 76]{minkowski}.
Take as an example the open question of inertia. We do not know why a body resists its acceleration.
But as the worldtube of an accelerating body is deformed (is not geodesic) it appears natural to assume
that a four-dimensional stress arises in the deformed worldtube of the body, which gives rise to a
restoring force that tries to restore the geodesic shape of the worldtube. This restoring force would
manifest itself as the inertial force\footnote{Calculations of the restoring force show that it does have the
form of the inertial force \cite{petkov}.} . A body whose worldtube is geodesic offers no resistance to its
motion because its worldtube is not deformed. Hence inertia may turn out to be another manifestation of
the four-dimensionality of the world.

The worldtube of a body falling in a gravitational field is geodesic and the body does not resist its fall.
However, the worldtube of a body, which is prevented from falling, is deformed and the restoring force
that arises in the body's worldtube has the same origin as in the case of an accelerating body but in this
case it would manifest itself as what is traditionally called the gravitational force. So, addressing the
issue of the reality of worldtubes sheds additional light on two questions: (i) Why is the gravitational force
considered to be inertial in general relativity \cite{rindler} (or, put another way, why ``there is no such
thing as the force of gravity'' in general relativity \cite{synge})? and (ii) Why are inertial and gravitational
forces (and masses) equivalent?


\end{document}